\documentclass[aps,prl,onecolumn,superscriptaddress,a4paper,12pt]{revtex4-2}
\usepackage{graphicx}
\usepackage{amsmath}
\usepackage{appendix}
\usepackage{color}

\begin{document}

\title{Efficient Quantum Simulation of Open Quantum System Dynamics on Noisy Quantum Computers}

\author{Shin Sun}
\affiliation{Department of Chemistry, National Taiwan University, Taipei City 106, Taiwan}

\author{Li-Chai Shih}
\affiliation{Department of Chemistry, National Taiwan University, Taipei City 106, Taiwan}

\author{Yuan-Chung Cheng}
\email[Email: ]{yuanchung@ntu.edu.tw}
\affiliation{Department of Chemistry, National Taiwan University, Taipei City 106, Taiwan}
\affiliation{Center for Quantum Science and Engineering, National Taiwan University, Taipei City 106, Taiwan}
\affiliation{Physics Division, National Center for Theoretical Sciences,
  Taipei City 106, Taiwan}

\date{\today}

\begin{abstract}
Quantum simulation represents the most promising quantum application
to demonstrate quantum advantage on near-term noisy intermediate-scale
quantum (NISQ) computers, yet available quantum simulation algorithms
are prone to errors and thus difficult to be realized. 
Herein, we propose a novel scheme to utilize intrinsic gate errors of
NISQ devices to enable controllable simulation of open quantum system
dynamics without ancillary qubits or explicit bath engineering, thus
turning unwanted quantum noises into useful quantum
resources. Specifically, we simulate energy transfer process in a
photosynthetic dimer system on IBM-Q cloud. By employing tailored
decoherence-inducing gates , we show that quantum dissipative dynamics
can be simulated efficiently across coherent-to-incoherent regimes
with results comparable to those of the numerically-exact classical
method. Moreover, we demonstrate a calibration routine that enables
consistent and predictive simulations of open-quantum system dynamics
in the intermediate coupling regime. This work provides a new
direction for quantum advantage in the NISQ era.
\end{abstract}

\maketitle

\section{Introduction}\label{sec:introduction}

Quantum computers promise revolutionary advantages in simulating quantum systems
highly important in physics and chemistry.\cite{nielsen2010quantum} The concept of quantum simulation
refers to utilizing quantum computers to simulate other quantum
systems of interest with resources much smaller than those required for the
same simulation in classical computers, and it could lead to
significant breakthroughs for scientific
research\cite{bacon2001universal, georgescu2014quantum}. 
Quantum simulation also represents one of the most critical and successful applications of
near-term quantum computers\cite{altman2021quantum}. However, although
various quantum simulation algorithms have been
proposed\cite{yung2014introduction,OMalley:2016dc,Colless:2018hp,cao2019quantum,McArdle:2020jl},
many of them have limited practical advantages over the classical
counterparts due to severe hardware constraints of current quantum devices.  
Contemporary quantum computers are prototypical
noisy intermediate scale quantum (NISQ)
devices\cite{preskill2018quantum,bharti2021noisy} as they are subjected to
various coherent and decoherent noises as well as limited in size and connectivity.

Since fault-tolerant universal quantum computation\cite{knill1998resilient,bravyi2018correcting} is unlikely
to be available in the near future, we are urged to find problems which are hard
to solve on classical computers but suitable for NISQ computers in
order to demonstrate quantum advantages for solving real-world
problems.
The simulation of open quantum systems is one of such problems
with significant applications, as dissipative dynamics of open
quantum systems\cite{breuer2002theory,weiss2012quantum} play crucial
roles in a broad range of physical and chemical phenomena of
complex systems, such as photosynthetic light harvesting\cite{Croce:2020hs,Yoneda:2020dh,Wang:2019ge},
and electron transfer in organic materials\cite{Rafiq:2019jw,Kim:2019ky}. Open quantum
systems also stand at the brink of quantum to classical transition,
which is of great importance for the fundamentals of quantum theory\cite{zurek2003decoherence}.
Hence, enabling efficient and accurate simulation of open quantum
system dynamics could lead to significant advances in both quantum 
science and technology.

Conventional methods for the simulation of open quantum system dynamics on
classical computers, such as the quantum master equation
approach\cite{breuer2002theory}, are based on certain weak-coupling
assumptions about the underlying system-bath interactions. When the
assumptions do not hold, non-perturbative treatments such as the
numerically exact hierarchical equation of motion (HEOM)
method\cite{Tanimura:1989ab, Tanimura:2006ab, Ishizaki:2009ab,Jin:2008ab} become
necessary in order to yield accurate results. However, these exact
methods on classical computers exhibit extremely steep scaling against
the accuracy and system size, which hinders our understanding of the
dynamical behaviors of complex open quantum systems.

On the other hand, there exist several proposals for simulating open
quantum system on quantum
computers.\cite{maniscalco2004simulating,chiuri2012linear,
	mostame2012quantum, potovcnik2018studying, wang2018efficient,
	trautmann2018trapped, maier2019environment, su2020quantum,
	garcia2020ibm,rost2020simulation} Most of these previous
      studies focused on theories and algorithms. For example,
Maniscalco \emph{et al.}\cite{maniscalco2004simulating} proposed to simulate the
quantum Brownian motion by engineering an artificial reservoir coupled
to a single trapped ion, and Aspuru-Guzik and coworkers\cite{mostame2012quantum} proposed an
superconducting analog open quantum simulator for Fenna-Matthews-Olson
complex, where the required system-bath interactions are introduced
though explicit engineering of qubit-resonator
couplings. Experimental realizations of quantum simulation of
open-quantum system dynamics on NISQ systems remain largely unexplored. Recently,
García-Pérez\cite{garcia2020ibm} \emph{et al.} 
simulated various decoherence channels on superconducting quantum
computers by explicitly realizing the non-unitary dynamics with 
ancillary qubits. 
Moreover, Wang \emph{et al.} \cite{wang2018efficient} utilized a NMR quantum
computer to simulate photosynthetic systems with the dephasing noises generated by
stochastically-modulated control fields. To our knowledge, the
four-qubit photosynthetic system they simulated remains the largest
open quantum system simulated on a
quantum computer. Nevertheless, due to the necessity of either
precise multi-qubit control or explicit system-bath engineering with
ancillary qubits, the resources overheads needed for these previously
published quantum simulation algorithms for open-quantum systems are
still formidable for nowadays NISQ quantum devices.

Let us recall Feynman in his famous quote \cite{feynman2018simulating}: 
\begin{quote}
	"Nature isn't classical, dammit, and if you want to make a simulation of nature, you'd better make it quantum mechanical, and by golly it's a wonderful problem, because it doesn't look so easy."
\end{quote}
In the same spirit, we might as well simulate dissipative quantum dynamics using intrinsic noises in quantum computers.
In this work, we propose to simulate open quantum system dynamics on
quantum computers by directly exploiting the errors generated on the
gate level in superconducting NISQ devices, which is an idea
that strongly differs from existing proposals. The
idea of utilizing decoherence and thermal effects in quantum computers to simulate quantum dynamics was
previously suggested by Lloyd and coworkers\cite{lloyd1996universal},
but the idea has never been realized. For most
quantum algorithms, quantum noises are undesirable and destructive.
Nevertheless, we recognize that the decoherence effects induced by quantum noises are
necessary and can be used as resources for simulations of open quantum
systems. Our approach thus does not require explicit engineering of the
control field, and no extra qubits for modeling the environment are
needed either.

In this paper, we demonstrate that gate errors on IBM-Q
cloud quantum computers can be controlled and utilized to simulate
excitation energy transfer (EET) dynamics in a molecular exciton dimer
system under tunable dissipative environments, and we compare the
simulation results with HEOM calculations to validate that they
correspond to dynamics induced by realistic system-bath
interactions. We also show that quantum computer can quantitatively
predict quantum dissipative dynamics at the intermediate system-bath
coupling regime that are hard to simulate accurately with classical
methods. Moreover, since the device instability for both qubit and
quantum gates is still high for NISQ
computers\cite{klimov2018fluctuations, burnett2019decoherence}, we
demonstrate designed calibration routines for noise strengths to
achieve consistent simulations. In summary, our approach shows the
possibility of utilizing NISQ devices as quantum noise generators as
well as embracing the noises to simulate physical systems. This opens
new directions for the simulation of open quantum system dynamics and
various stochastic systems such as in biology, finance, or
cryptography.

\section{Quantum Simulation of Coherent Energy Transfer}

\subsection{Exciton Model}
To demonstrate the simulation of open quantum system dynamics on noisy
quantum computers, we adopt a simple exciton
model that describes EET in photosynthetic
systems. In the model, physical systems are represented as
different monomers called "sites", which constitute the whole complex, 
and two quantum states, the ground state and the excited state, are
considered for each site. This two-level property makes
them mapped naturally to qubits, and thus ideal target systems to be
simulated on quantum computers. 
The Hamiltonian describing an excitonic system with N sites in the second-quantized form is 

\begin{align} \label{exciton_def}
	H &= \sum_{i=1}^{N}\epsilon_i {a_i}^\dagger a_i + \sum_{i\neq j=1}^{N}J_{ij}{a_i}^\dagger a_j
\end{align}
where ${a_i}^\dagger(a_i)$ is the creation (annihilation) operator acting on site $i$. 
It is convenient to introduce the state $|0\rangle$, which corresponds to the state with all sites in the ground state, and $|i\rangle={a_i}^\dagger|0\rangle$, which corresponds to the state with a single excitation at site $i$ while other sites are in the ground state.
The diagonal matrix element $\epsilon_i$ is the site energy of $|i\rangle$, and the off-diagonal
term $J_{ij}$ corresponds to excitonic coupling between $|i\rangle$
and $|j\rangle$. For simplicity, we consider a symmetric dimer system
($N=2$, $\epsilon_1 = \epsilon_2 = 0$) with excitonic coupling
$J_{12} = J_0$. This model also corresponds to the spin-boson problem 
or the Heisenberg XY model for magnetism\cite{lieb1961two}. In this work, we specifically
focus on the dynamics in the one-exciton manifold for EET.

\subsection{Quantum Circuit and Simulated Coherent Dynamics}

\begin{figure}
        \centering 
	\includegraphics[width=0.95\columnwidth]{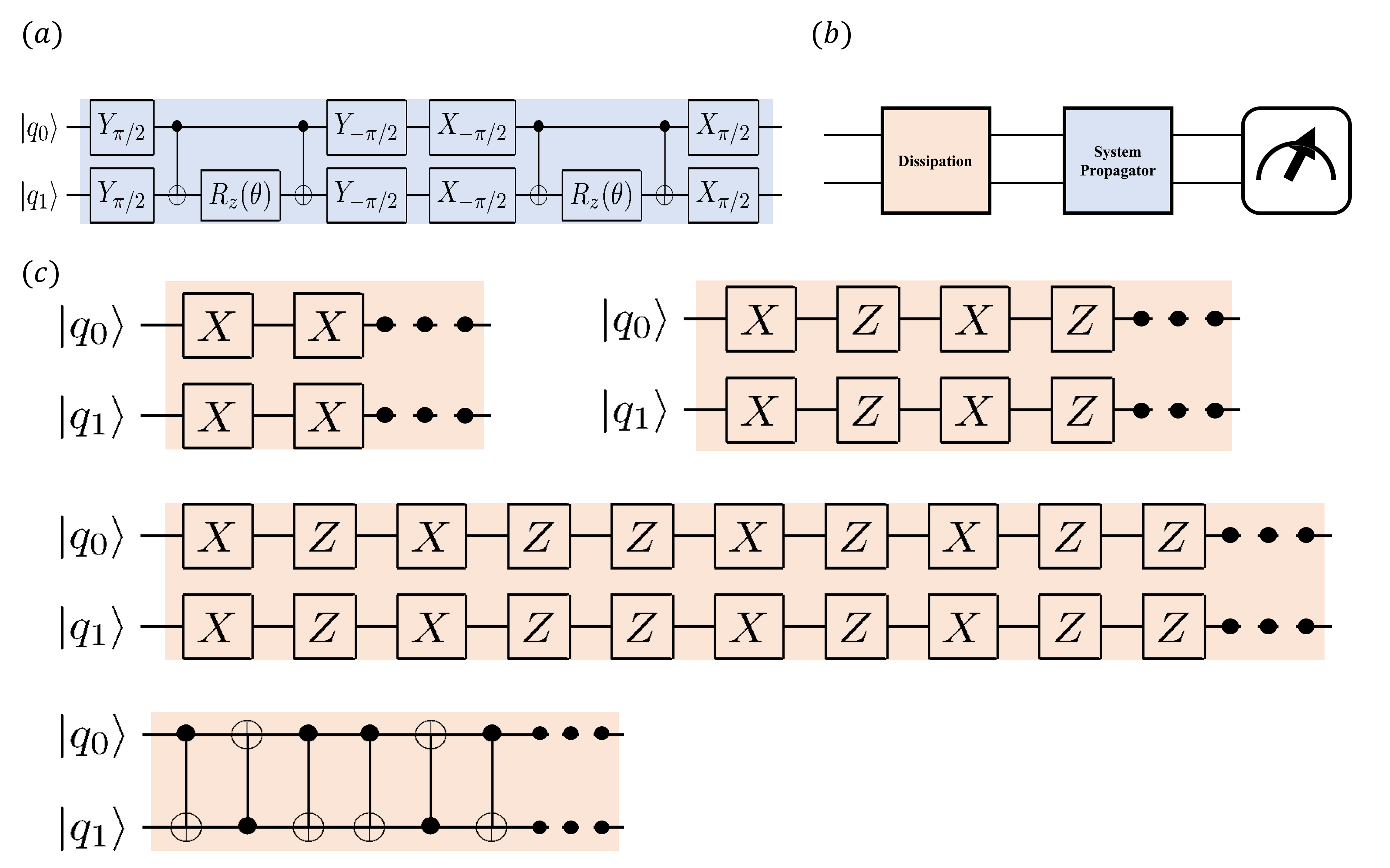}
	\caption{\textbf{Quantum circuits for the
            dissipative quantum system simulation.}  (a) The circuit
          implementing the time-evolution propagator for the unbiased exciton dimer in
          our experiments. (b) Illustration of the simulation scheme. The full simulation circuit first
          undergoes dissipative operation via different decoherence-inducing gates that introduce decoherence
          into the simulated system, and followed by the 
          propagator of the system Hamiltonian. (c)
          Various decoherence-inducing gate sequences used for the
          dissipation circuit.
\label{fig:quantum_circuit}}
\end{figure}

To represent and propagate the exciton dynamics efficiently on quantum
computers, we encode the exciton occupancies of the dimer system in the computational
basis of two qubits. The Hamiltonian of the system is mapped to Pauli
operators acting on the qubit Hilbert space via a Jordan-Wigner type
transformation, and the details of the encoding circuit are
described in Appendix~\ref{sec:exp_methods}. Thus, the quantum circuit which simulates 
the system propagator $e^{-iHt}$ ($\hbar=1$) can be constructed as shown in 
Fig.~\ref{fig:quantum_circuit}~(a), where $\theta$ corresponds to the simulation time.
Note that the resulting circuit for the propagator is
exact in the symmetric dimer case due to the commutativity of the two
terms in the system Hamiltonian. More generally, one has to build the
propagator though Trotterization or other efficient state propagation
schemes\cite{low2017optimal} for a general biased excitonic system. To
retrieve the dynamical population information, we perform projective
measurements in the qubit computational basis, and renormalize 
the probabilities within the one-exciton manifold (see Methods) to
compensate for the leakage errors. 

\begin{figure}
        \centering 
	\includegraphics[width=0.5\columnwidth]{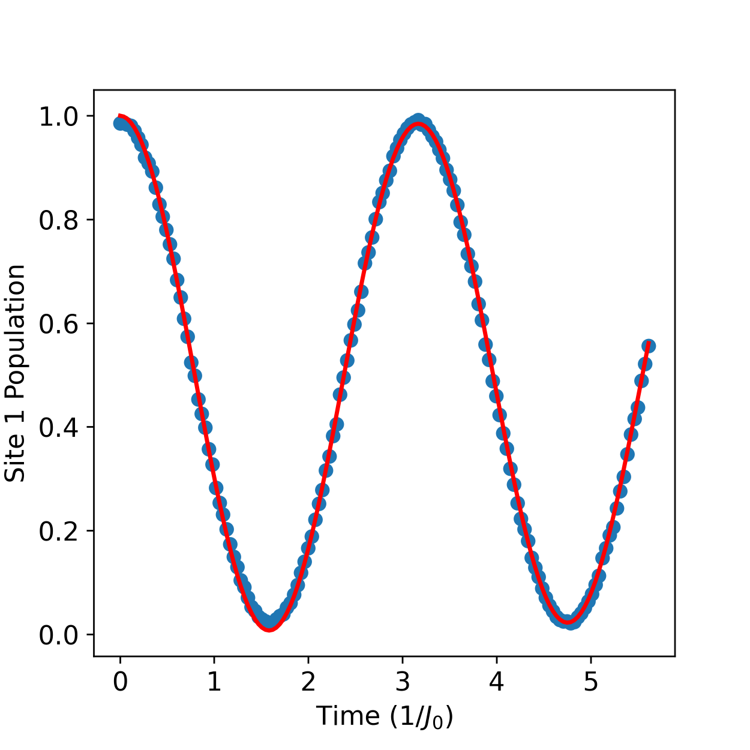}
	\caption{\textbf{Simulated coherent dynamics of the dimer system.} 
		  Population on site 1 as a function of time is depicted when the 
		  system is initially prepared as $|\psi(0)\rangle=|1\rangle$.
		  The simulation result obtained on ibmq\_bogota (blue dots) and a cosine
		  function (red line) are both shown.
\label{fig:pop_rabi}}
\end{figure}

We experimentally simulated the time evolution of the dimer system with initial population on site 1 using
the quantum circuit depicted in Fig.~\ref{fig:quantum_circuit}(a) on the
IBM-Q superconducting quantum computers\cite{IBMQuantum}.
Figure~\ref{fig:pop_rabi} depicts the simulated population dynamics. The results show a perfect
coherent Rabi oscillation ("quantum beating"). This validates the mapping and confirms that our renormalization procedure to account for the leakage error does not affect the coherence in the one-exciton manifold.

\section{Tailored Decoherence-inducing Gates}

\subsection{Decoherence Engineering by Prepending Identity Gate Sequences}

To simulate realistic energy transfer dynamics on IBM-Q systems, we
aim to utilize intrinsic imperfections in the quantum gates
to model a dissipative environment in a controllable manner. To this
end, we prepend a dissipation part to the simulation circuit
by a number of gate sequences that ideally contract to the identity gate (Fig~\ref{fig:quantum_circuit}(b)). Due to the gate errors, the identity gate sequences will introduce quantum noise into the simulation on NISQ devices. We thus refer to them as "decoherence-inducing" gates in this work. The system propagator then changes the frame of the system from the interaction picture to Schr\"{o}dinger picture
at a given simulation time. Finally, projective measurements are performed to monitor site populations.

To tune the strength of system-bath couplings, we control the number of decoherence-inducing gates applied per unit
simulation time. Therefore, we define a damping coefficient $d = \frac{N_I}{t / \Delta T_D}$, where 
$t$ is the total simulation time, $\Delta T_D$ is the decoherence period, and $N_I$ is the number of decoherence-inducing
gates applied. The damping coefficient can then be adjusted 
in the quantum simulation to realize different system-bath coupling strengths.

Regarding decoherence-inducing gates, we examined four types of identity gate sequences as shown in Fig.~\ref{fig:quantum_circuit}(c), including
1.$(X)^2$ 2.$(XZ)^2$ 3.$(XZXZZ)^2$ 4.$(SWAP)^2$. Gate sequences 1-3 act on a single qubit, 
while gate 4 is a non-local gate. These gate sequences are chosen
based on a series of preliminary studies on noisy qubit dynamics
induced by various quantum gates available on the IBM-Q systems (see Sec. S1-S5 in the SI). We noted that because the virtual-Z gate implementation on IBMQ systems
\cite{mckay2017efficient} corresponds to a mere frame
shift of the subsequent pulses, it does not introduce a real time
elapse in the simulation (i.e. no pulse is applied). Hence, Z gates
cannot introduce gate noises (Fig. S6). We therefore
choose X gate as our primary decoherence-inducing single-qubit gate. In the following, we systematically
analyze the dissipative dynamics induced by the four types of gate sequences, and details of the implementation
and parameters used in this work are given in the Methods section.

\subsection{Dissipative Dynamics Induced by Single-qubit Gate Sequences}
\begin{figure}
        \centering 
	\includegraphics[width=0.9\columnwidth]{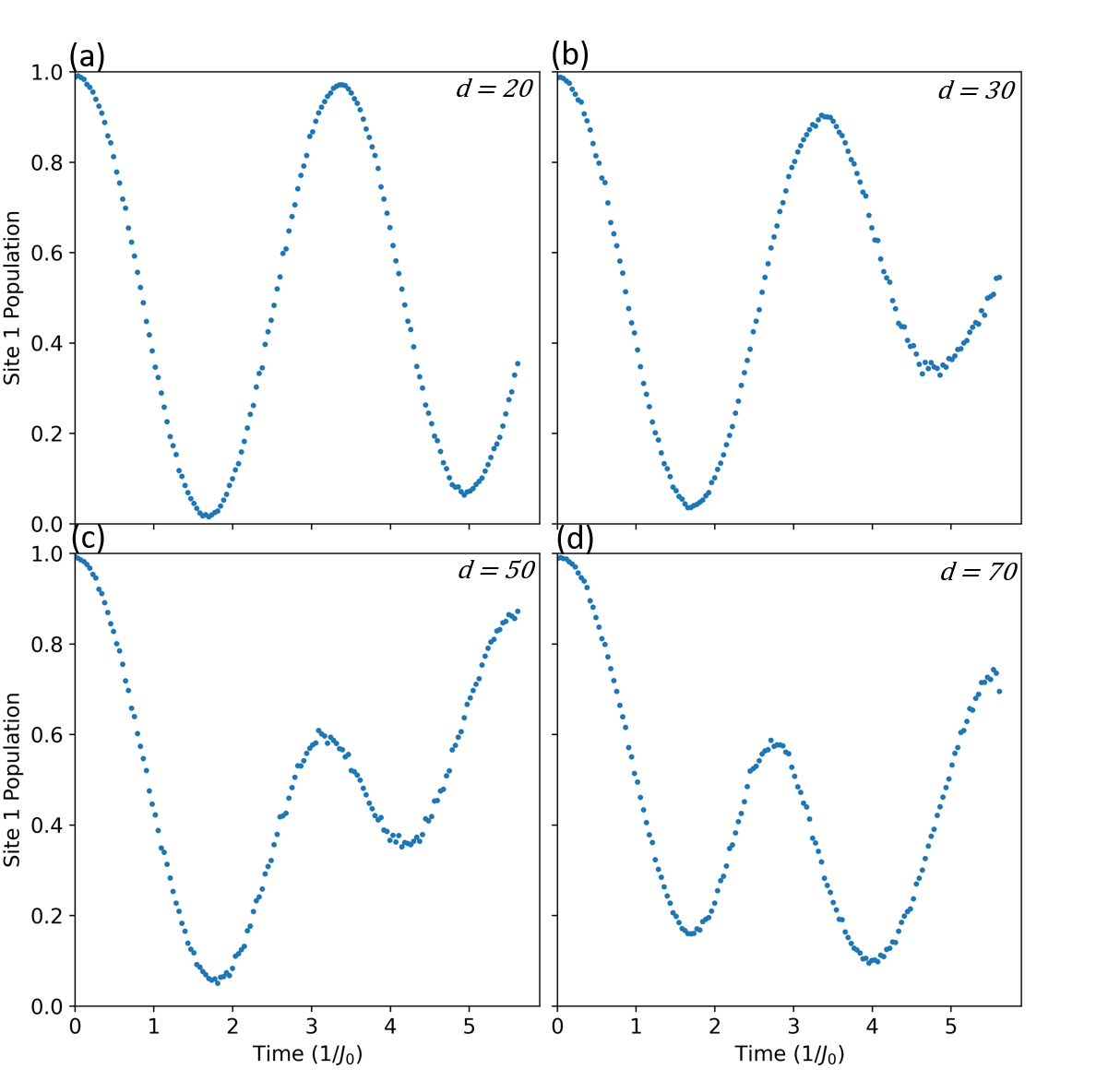}
	\caption{\textbf{Simulated dissipative dynamics induced by $(X)^2$.} 
		  We plot the population on site 1 (blue dots) as a
          function of time at 4 different damping coefficients. The
          decoherence effects are induced by applying the
          $(X)^2$ gates. (a) $d=20$, (b) $d=30$, (c) $d=50$, and (d) $d=70$.
          Experiments were performed on ibmq\_bogota. 
\label{fig:pop_XXdissipative}}
\end{figure}

Three types of single-qubit gate sequences are examined in this work.
We first employ $(X)^2$ gates to simulate EET dynamics of the model dimer system. Figure~\ref{fig:pop_XXdissipative}
shows the simulated population dynamics at four different damping coefficients. At small damping coefficients ($d=20$ and $d=30$), the decoherence of the quantum 
beating is apparent and becomes more pronounced as $d$ increases, which are in line with our proposal.
However, at larger damping coefficients ($d=50$ and $d=70$), the population dynamics exhibit unphysical multifrequency behavior and
large population revival; thus, the simulated dynamics cannot reach
equilibrium and do not correspond to physical dissipative
dynamics. These results indicate that $(X)^2$ does not introduce
realistic decoherence effects into the EET dynamics and thus cannot be
used as a proper decoherence-inducing gate.

To elucidate the errors accompanying the $(X)^2$ gate sequence, we
performed state tomography after applying
$(X)^2$ gates on a single qubit initialized at a given each basis
state to analyze the 
dynamics induced (see Sec. S1-S4 in the SI). Figure S3 indicates that 
$(X)^2$ error consists of significant over-rotation of the X gate, which 
leads to a systematic drift in the dynamics and thus produces non-physical
results in the EET simulation. In other words, applying $(X)^2$ gates results
in the simulation of an effective Hamiltonian that is different from the target
Hamiltonian. The issue could have been resolved by adding an
offset to each X rotation to counteract the over-rotation, but the
downside is that the over-rotation angle is then needed for each
experiment beforehand, which is not practical given that the stabilities
of the available quantum computers are still less than perfect.  

In order to eliminate the drift in the X gate error and obtain
more consistent random noises, we employ echo-type
equivalents to the $(X)^2$ gate, the $(XZ)^2$ gate, and the
corresponding higher order $(XZXZZ)^2$ gate as the
decoherence-inducing gates. The
design of these gates is related to the compensating pulse sequence
techniques\cite{merrill2014progress} (see Sec. S5 in the SI), which dynamically correct
the X over-rotation. Figures S7 and S8 show the time-evolution of a single qubit induced by the
$(XZ)^2$ and $(XZXZZ)^2$ gates, respectively. Due to the phase
error, the $(XZ)^2$ does not remove the over-rotation completely. 
Nevertheless, the $(XZXZZ)^2$ sequence clearly induces a depolarization-like random noise, and it can thus
be used to model a physical decoherence process.

It is interesting to note that the idea of using controlled coupling
of a quantum system to its environment to engineer open-system
dynamics has been realized previously. For instance, Barreiro \emph{et
  al.}\cite{barreiro2011open} 
utilized dissipative quantum systems on trapped-ion
architectures by combining multi-qubit gates and optical pumping
techniques to prepare highly entangled quantum states. 
More recently, Rost \emph{et al.} \cite{rost2020simulation} 
proposed a scheme for simulating open quantum systems by using
intrinsic qubit noise. In their work, the decoherence effects are
introduced to the quantum simulation via the "idle" operation, which
exhibits minimal overhead for simulating noises, but then the
achievable bath conditions are solely determined 
by the intrinsic qubit properties. Our study indicate that the
intrinsic qubit dissipation and gate errors could be complicated and
can not be used for controlled open-system
dynamics engineering. Therefore, in our proposed scheme, we shift the
attention from imperfections in qubits to pulses, and this allows us
to control the system-bath couplings in a much more flexible manner of pulse
engineering. Consequently, our work significantly differs from
previous methods in the implementation of the tailored
decoherence-inducing gate sequences. 

\begin{figure}
        \centering 
	\includegraphics[width=0.9\columnwidth]{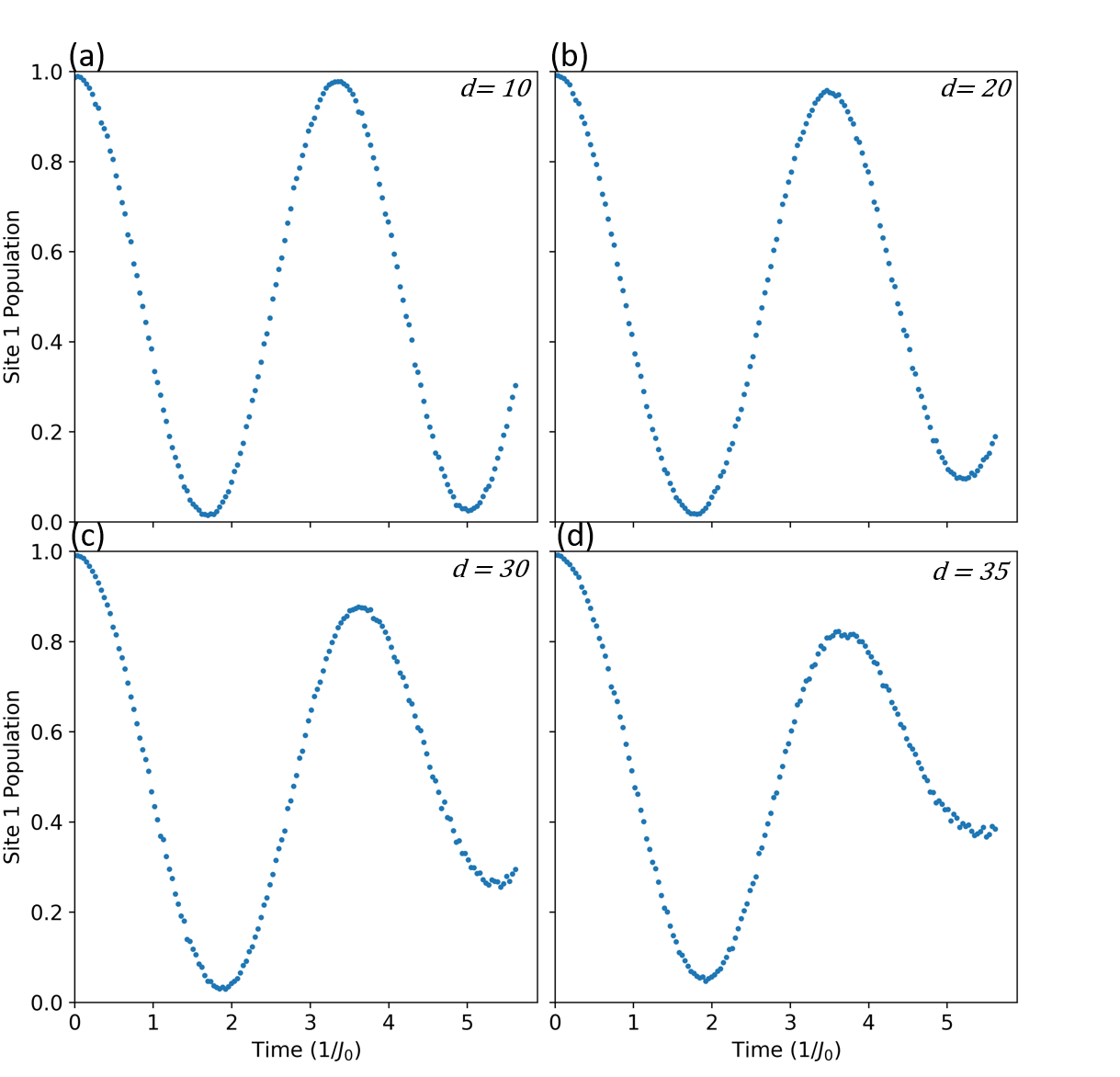}
	\caption{\textbf{Simulated dissipative dynamics 
			 induced by $(XZXZZ)^2$.} 
		  We plot the population dynamics on site 1 as a
          function of time under 4 different damping coefficients. The
          decoherence effects are induced by applying the
          $(XZXZZ)^2$ gates. (a) $d=10$, (b) $d=20$, (c) $d=30$, and (d) $d=35$.
          Experiments were performed on ibmq\_bogota. 
\label{fig:pop_XZXZZ}}
\end{figure}

In the results shown in Fig.~\ref{fig:pop_XZXZZ}, we employed
$(XZXZZ)^2$ as the decoherence-inducing gate, and assess the population dynamics for the EET process 
under different damping coefficients. The decoherence effect can be seen in the damping of the quantum beating,
and the decay rate increases with the increase of damping coefficient consistently. This confirms that the $(XZXZZ)^2$
can be used as a decoherence-inducing gate sequence to introduce quantum noises into the simulation. However, even when the damping coefficient is large ($d=35$, where the circuit depth is closed to 
the device limit), the population dynamics still remain coherent, and this restricts us from exploring the EET dynamics in the incoherent regime. Furthermore, the maximum time duration of the simulation is limited by the deepest circuit depth available on the NISQ devices. As $d$ increases, the number of gates needed to perform simulation in a unit time also increases, which limits the available simulation time.
Therefore, a more general decoherence-inducing gate sequence is desirable.

\subsection{Dissipative Dynamics under $(SWAP)^2$ Gate Sequences}
\begin{figure}
        \centering 
	\includegraphics[width=0.9\columnwidth]{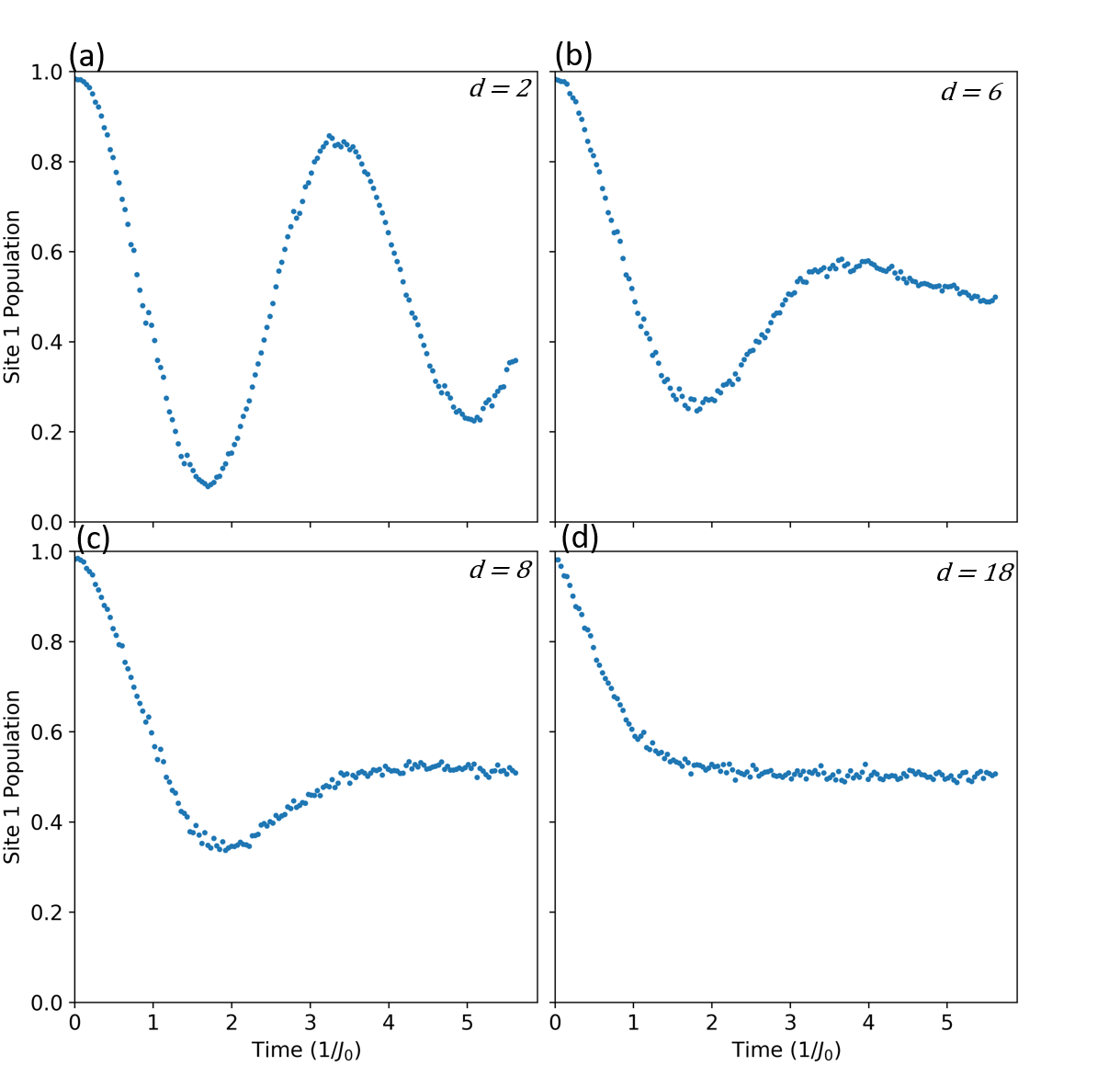}
	\caption{\textbf{Simulated dissipative dynamics 
			induced by $(SWAP)^2$.} 
		  We plot the population dynamics on site 1 as a
          function of time under 4 different damping coefficients. The
          decoherence effects are induced by applying the
          two-qubit $(SWAP)^2$ gates. (a) $d=2$, (b) $d=6$, (c) $d=8$, and (d) $d=18$.
          Experiments were performed on
          ibmq\_paris. 
\label{fig:pop_SWAP}}
\end{figure}

In addition to single-qubit gates, two-qubit gates could also be utilized to construct decoherence-inducing gate sequences.
We discovered that the non-local $(SWAP)^2$ gates on IBM-Q systems can introduce consistent
decoherence effects into the model system and drive the system towards
equilibrium efficiently. In
Fig.~\ref{fig:pop_SWAP}, the population dynamics under $(SWAP)^2$ with
different damping coefficients for the model system are depicted. The
decay of the quantum beating is much faster compared to results shown in Fig.~\ref{fig:pop_XZXZZ}, suggesting that the
noise introduced by the $(SWAP)^2$ gate is stronger than that introduced by the $(XZXZZ)^2$ gate. The system even
exhibits incoherent transfer dynamics at a large $d$ (Fig.~\ref{fig:pop_SWAP}(d)), which enable us to study the dissipative dynamics in the 
strong coupling regime. Since the dissipative
dynamics under $(SWAP)^2$ gates are consistent throughout
coherent to incoherent regime, we quantitatively analyze and
compare them to realistic open quantum system dynamics. The
possibility of combining different type of gate sequences or other
pulse-level techniques to achieve finer control over system-bath
couplings is left for future research.

\begin{figure}
	\centering 
	\includegraphics[width=0.5\columnwidth]{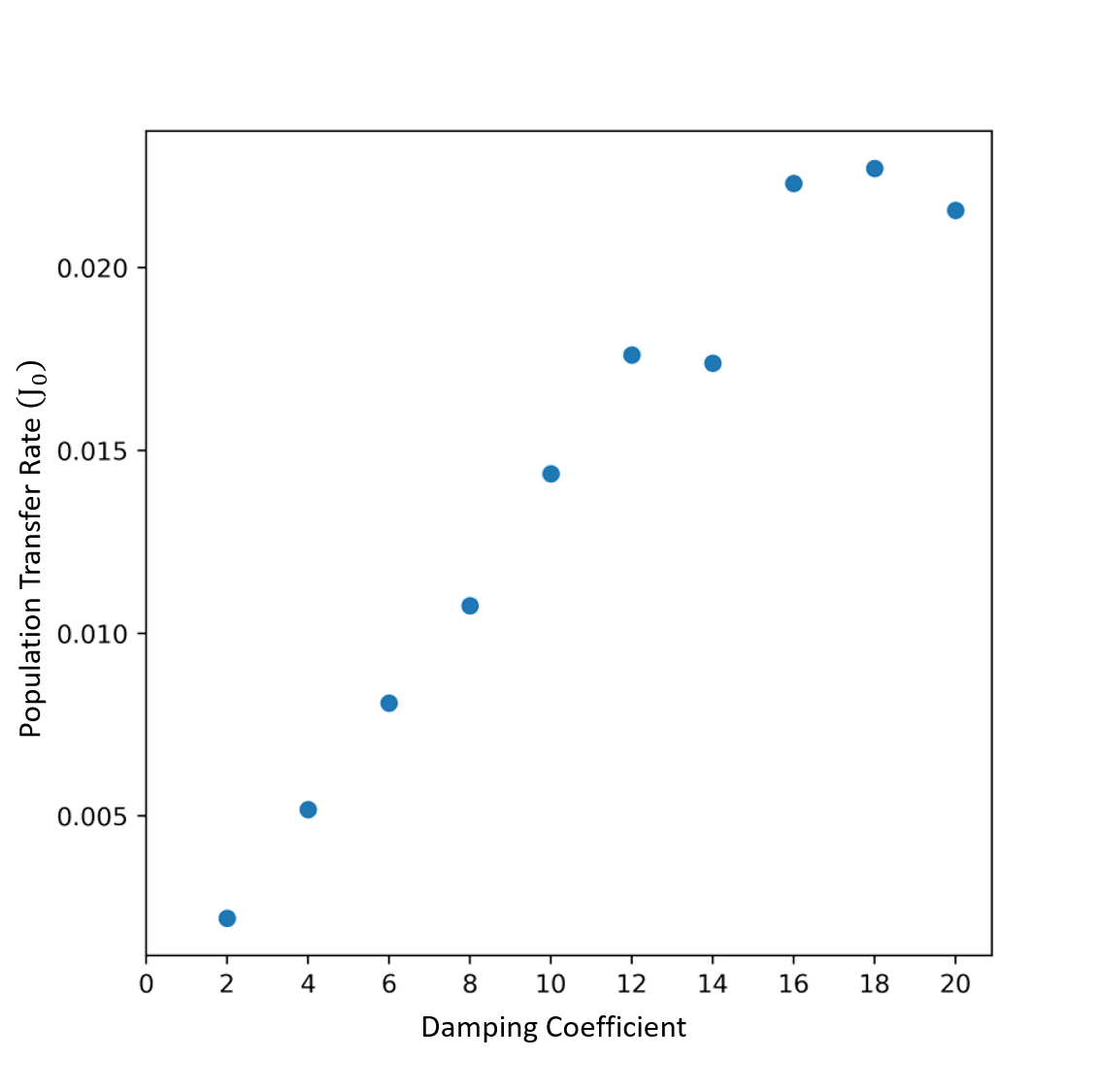}
	\caption{\textbf{Simulated population transfer rate as a function of
			damping coefficient.} 
			Blue dots depict population transfer rate from $|1\rangle$ to $|2\rangle$ extracted from quantum simulation using 
			$(SWAP)^2$ as decoherence-inducing gate sequence at different damping coefficients. Experiments were performed on
			ibmq\_paris
		\label{fig:pop_rate_D}}
\end{figure}

Furthermore, we extract the population transfer rate by fitting the
dynamics at each damping coefficient with an exponential-cosine 
function, and plot the rates against the damping coefficients (Fig.~\ref{fig:pop_rate_D}).
The relation between decay rate
and damping coefficient varies under different system-bath coupling strengths. At weak system-bath couplings (small damping coefficients), the decay
rate approximately increases linearly with respect to the damping coefficient ($d=0$ to
$d=10$); however, as the damping coefficient further increases ($d>10$), the
rate exhibits a turnover behavior. It is interesting and important to note that the
relation is not monotonic, suggesting an optimal transfer rate at the intermediate coupling regime in accordance with previous EET studies\cite{Ishizaki:2009ab,Mohseni:2008p67347,Chang:2012il}. The behavior also suggests that our
simulation results are non-trivial in the sense that they cannot be 
reproduced by simple Redfield-type theories. Note that the
computational costs in terms of number of qubits and circuit depth
needed in our proposed scheme are linear with respect to the number of
sites ($N$) and simulation time ($t$), respectively. Moreover, the
dissipative part can be introduced by using at most two-qubit gates
that can be executed in parallel on nowadays superconducting quantum
computers. Thus, the proposed scheme should be able to be easily
scaled up to simulate open quantum system dynamics for multiple-site
systems.

\section{Validity of the Method}

\subsection{Comparison with Numerically Exact HEOM Simulations}

To validate that our simulation based on $(SWAP)^2$ decoherence-inducing gate represents dynamics of a realistic open
quantum system, we compare our results to those obtained by classical computation based on a microscopic 
modeling of system-bath interactions.
To this end, we fit our quantum simulation data to those obtained from the hierarchical equation of
motion using the Parallel Hierarchy Equations of Motion
Integrator program\cite{strumpfer2012open} on classical computers. 
The HEOM method employs a Drude-Lorentz type spectral density to model system-bath
interactions:
\begin{align} \label{eq:spectral_density}
	J(\omega) = & \frac{\lambda}{2}\frac{\gamma\omega}{\gamma^2 +
                      \omega^2} 
\end{align}
where $\lambda$ is the reorganization energy, representing the
system-bath coupling strength, and $\gamma$ is the cut-off frequency
that corresponds to the inverse bath relaxation time scale. HEOM has
been recognized as a highly accurate method for EET dynamics; however,
the full HEOM method requires solving a system of non-polynomial
number of differential equations. As a result, HEOM calculations need
truncation at a certain hierarchical level and exhibit extremely stiff
computational cost against system size, making it best suited for
small-size systems. 

\begin{figure}
        \centering 
	\includegraphics[width=\columnwidth]{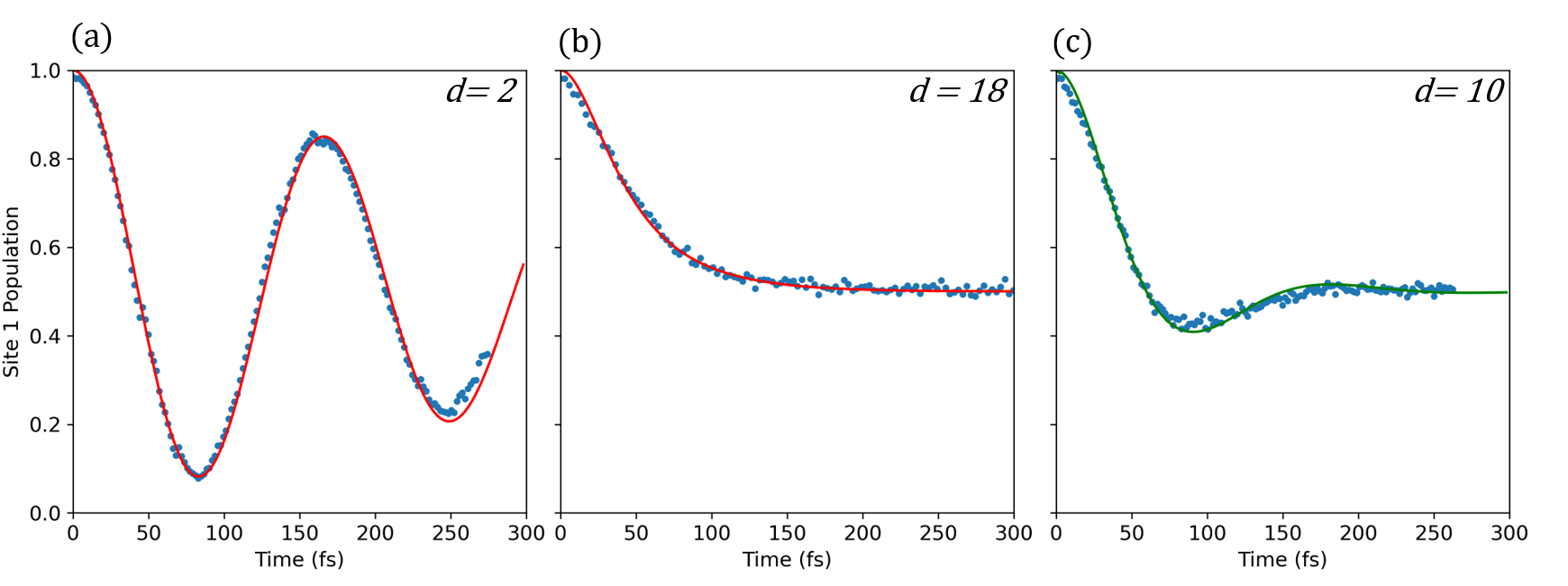}
	\caption{\textbf{HEOM analysis of the simulated 
            dynamics induced by $(SWAP)^2$ gates.} The simulated
          population dynamics at three system-bath coupling parameters
          are shown: (a) weak coupling ($d=2$), (b) strong coupling
          ($d=18$), and (c) intermediate coupling ($d=10$). The weak-
          and strong-coupling results are fitted individually with
          HEOM simulations, of which the results are shown as red
          lines and the fitting parameters are listed in Table
          ~\ref{tab:table1}. With the fitted data, we
          assume a linear relationship between the reorganization energy and
          damping coefficient, and use quantum simulation to predict
          the dynamics at $\lambda=120\ \textrm{cm}^{-1}$. The
          predicted dynamics and the corresponding HEOM dynamics
          are shown as dots and the green line, respectively, in
          (c). 
\label{fig:heom_fit}}
\end{figure}

We choose to fit HEOM dynamics to the dynamics simulated using the
$(SWAP)^2$ decoherence-inducing gates at two cases, a weak coupling
case with $d=2$ and a strong coupling case with $d=18$. For the HEOM
model parameters adopted in the fitting, we assume that each site in
the dimer is coupled to identical and independent baths, and fix
$J_0=100\ cm^{-1}$, $\gamma=100\ ps^{-1}$, and the temperature at 300
K. We only vary $\lambda$ in the HEOM model to reproduce the simulated
dynamics using $(SWAP)^2$ gates on IBM-Q systems. In
Fig.~\ref{fig:heom_fit}(a) and Fig.~\ref{fig:heom_fit}(b), we compare
the population dynamics simulated on the IBMQ devices 
with the best HEOM fit for the $d=2$ and $d=18$ cases,
respectively. The optimal HEOM parameters are listed in
Table. \ref{tab:table1}. The results demonstrate that both coherent
and incoherent EET dynamics simulated by the $(SWAP)^2$
decoherence-inducing gates on IBMQ devices can be quantitatively
reproduced by microscopic HEOM calculations. Evidently, our proposed
scheme can simulate realistic dissipative quantum dynamics 
with controllable system-bath coupling strengths.

Furthermore, the
excellent agreement between dissipative dynamics induced by $(SWAP)^2$
gates and HEOM simulation of a microscopic model system provides
crucial insights into the characteristics of the errors induced by the
$(SWAP)^2$ gates. The Drude-Lorentz bath model adopted in the HEOM
method represents an overdamped Brownian bath with Gaussian random
fluctuations, and it should not be too surprising that the gate errors
in IBM-Q systems could be quantitatively described by such a random
process. In this regard, the fit to HEOM dynamics provides a
consistent framework to extract error characteristics in quantum gates
implemented in a quantum computer, as shown in the bath parameters
listed in Table.~\ref{tab:table1}. Such characteristics could be
dependent on the type of decoherence-introducing gates as well as the
hardware, and could have significant implications in optimizations of
quantum algorithms and even designs of quantum error-correction
schemes. This direction of research is a work in progress and will be
reported in another paper.

Knowing that the simulation scheme has worked, we then investigate if we could use the scheme to predict the dissipative
dynamics at other system-bath coupling strengths by controlling the
damping coefficient. To this end, we assume a linear relationship between damping coefficient and bath reorganization energy in the microscopic model. 
According to Table ~\ref{tab:table1}, we can determine the required damping coefficient used in the quantum simulation for simulating dynamics with a given reorganization energy.
For instance, a model system with $\lambda=120\ cm^{-1}$ should correspond to the quantum simulation with $d=10$. In Fig.~\ref{fig:heom_fit}(c),
we plot the simulated dynamics at $d=10$ and the HEOM result with $\lambda=120\ cm^{-1}$. Note that there are no free parameters in both simulation shown here. The excellent agreement between them
provides strong numerical evidence to support that the proposed scheme for dissipative quantum dynamics simulation can be used as a predictive tool, especially in the difficult intermediate coupling regime.

\begin{table}
	\caption{\label{tab:table1} The fitted HEOM parameters for the
          dissipative dynamics shown in
          Fig.~\ref{fig:heom_fit}.}
	\begin{ruledtabular}
		\begin{tabular}{lcc}
			& Underdamped&  Overdamped\\
			\hline
			Damping coefficient & $2$ & $18$ \\
			Excitonic coupling & $100\ cm^{-1}$ & $100\ cm^{-1}$\\
			Cutoff frequency & $100\ ps^{-1}$ & $100\ ps^{-1}$\\
			Reorganization energy & $15\ cm^{-1}$ & $227\ cm^{-1}$\\
			Hierarchy truncation & 8 & 8\\
			Temperature & 300 K & 300 K
		\end{tabular}
	\end{ruledtabular}
\end{table}

\subsection{Temporal Stability and Calibration Scheme}

\begin{figure}
        \centering 
	\includegraphics[width=0.8\columnwidth]{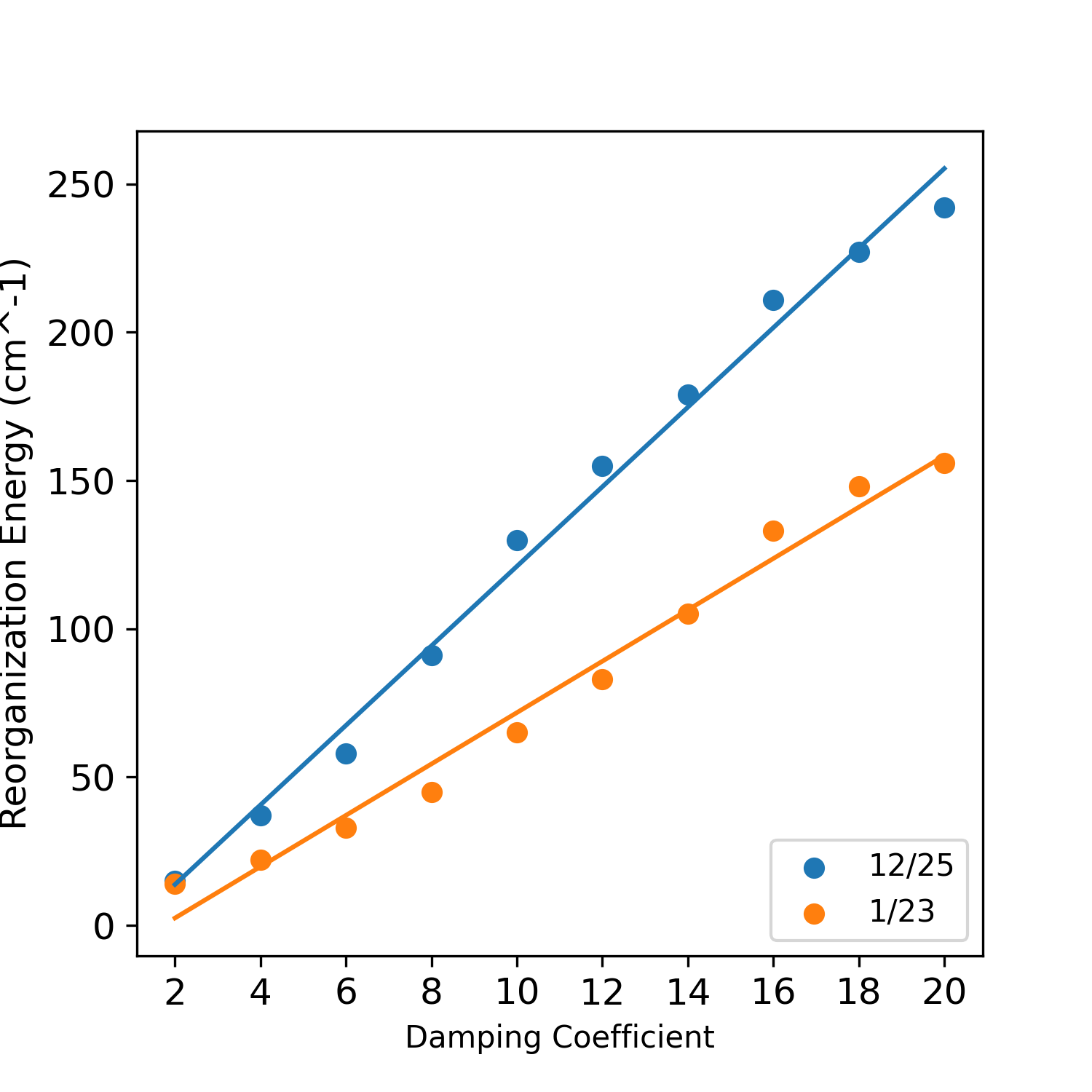}
	\caption{\textbf{Calibration curves on different days.} We
          plot the reorganization energy obtained from HEOM fitting
          against the damping coefficient used to generate quantum
          simulation data for experiments performed on different
          days. Blue dots denote results obtained on
          2020/12/25, and yellow dots are results obtained on
          2021/1/23. All
          experiments were performed on ibmq\_paris. 
	}
\label{fig:calibration}
\end{figure}

 During our study, we also found that the gate noises on IBM-Q superconducting devices
 in general could vary on a daily basis (see Sec. S6 in the SI). Therefore, it is crucial that we assess
 the stability of our scheme on different days. We thus performed on different days a series of simulations of the EET dynamics using $(SWAP)^2$ gates with different damping coefficients, and for each damping coefficient we
 fit the simulated population dynamics with HEOM calculations to
 analyze the relationship between the fitted reorganization energies
 and damping coefficients. In Fig.~\ref{fig:calibration}, we plot the
 calibration curves of the fitted reorganization energies against the
 damping coefficients for two separated sets of simulations executed on different days. We observe that the two curves remain linear, yet their slopes are
 different, which indicates that the device shows different level of
 noise strength in different days, and it could affect our simulation
 significantly. 

Nevertheless, the linear relationship for the calibration curve on
each day suggests that it is possible to run a small number of
experiments to calibrate the noise strengths. Ideally, two experiments, one for a weak coupling case and the other
for a strong coupling case, could be carried out and fitted to
classical simulations to determine the noise characteristics. Note
that accurate
methods to simulate the dynamics in these two extremes are not as costly as simulating
dynamics at the intermediate coupling regime \cite{breuer2002theory}, and therefore the HEOM method is not strictly required if a large system is under study.
The relation can then be used to simulate dissipative
dynamics under other system-bath coupling strengths by
controlling only the damping coefficient on quantum computers.
The fluctuation of noise strengths for the
 single-qubit decoherence-inducing gates is discussed in Sec. S6 in
 the SI.
Other methods to efficiently calibrate the noise strengths and, possibly, stabilize them are
left for further research.  

\section{Concluding Remarks}
In this work, we have successfully simulated the EET dynamics for a symmetric dimer system under different damping
regimes using only the intrinsic gate noises of the IBM-Q systems.
Our approach requires neither additional ancillary qubits to represent the
environment nor explicit bath engineering on the hardware
level. Significantly, we show that by designing gate sequences to produce consistent random noises, the simulation scheme can yield results that are in excellent agreement with HEOM simulations of microscopic EET models. Moreover, although current superconducting quantum devices does not
offer enough stability over time for consistent quantitative
simulation of dissipative quantum dynamics, we show that a simple calibration
scheme can be applied to turn NISQ devices into a controllable simulator for open
quantum system dynamics to successfully predict the EET dynamics
in the intermediate coupling regime.  

In principle, the decoherence-inducing gate method proposed here only
needs N qubits to simulate dissipative dynamics of a N-site system,
and with superconducting quantum computer with more than 100 qubits
readily available in the near future, it is possible to use our method
to simulate large and complex excitonic system such as the
photosynthetic photosystem I ($\sim$ 100 sites) and photosystem II
supercomplex ($\sim$ 300 sites), as well as conjugated polymers and a
matrix of small-molecule chromophores used in organic solar
cells. Realization of such simulations could significantly advance our
understanding of elementary photophysical processes on these crucial
systems. On the other hand, accurate classical simulation methods such
as the HEOM approach demand significantly more computational cost and
can not be easily applied to systems \textgreater 100 sites. Generally
speaking, with extremely efficient coding and parallel algorithmic
optimization, the HEOM method could be applied to simulate a system
with about 50 sites, but further scaling up would be
formidable. Therefore, we present a quantum simulation algorithm that
is highly likely to demonstrate clear quantum advantage in the near
future. Note that the advantages comes naturally from using NISQ
devices as a quantum noise generator, and our study indicates that the
task is not trivial as nowadays NISQ devices could exhibit coherent
noises that are not described by simple stochastic processes. The key
insight in this work is that by choosing and designing specific
compensating pulse sequences, purely random gate errors can be
realized and controlled. Hence, intrinsic gate errors could and must
be engineered to become a type of quantum resources. 

We believe that the major roadblock of the scheme proposed in this
work remains the limited circuit depth available on nowadays NISQ
devices. For example,


We also noted that there exists a slight discrepancy between simulated dynamics 
and HEOM fitting at long time for the dynamics shown in
Fig.~\ref{fig:heom_fit}(a), which can be attributed to accumulated errors in the deep
quantum circuit needed to simulate long time dynamics. One possible method to remedy this inaccuracy in the long time dynamics is to use the transfer tensor
algorithm\cite{cerrillo2014non} to tease out dynamical correlations in quantum simulation at
the initial stage, and propagate the dynamic classically to an
arbitrary long time. Such hybrid quantum-classical approach would only
need shallow quantum simulations and could
make it possible to overcome the circuit depth problem. Of course, traditional transfer tensor algorithm needs
dynamical map at each time step, and building up dynamical maps
requires process tomography\cite{mohseni2008quantum} on quantum
computers. That makes the scaling not ideal for large system. Thus,
we envision that a more compact scheme to obtain dynamical maps based
on partial information of the underlying dynamics like compressed
sensing techniques\cite{gross2010quantum, shabani2011efficient} should
be used if one is to extend the initial trajectories of the
dissipative simulation of dynamics based on our scheme.   

Our research also suggests a number of possible improvements. First
of all, we only explored a controllable bath parameters that is  
the reorganization energies $\lambda$. It is important that a systematic
procedure to adjust the cutoff frequencies $\gamma$ can be found in order 
to generate baths with any desirable spectral properties. Presumably, dynamics with different $\gamma$ can be simulated
by designing different decoherence-inducing gates or by choosing a different unit-less energy scaling. Furthermore, additional research directions including detailed characterization of the effects of various decoherence-inducing gates, 
finer control over bath parameters using different gates or pulse level control
(like pulse stretching), scheme to provide classically hard-to simulate quantum noises, 
implementing finite temperature effect for biased exciton systems, and
extending the simulations scheme to study larger systems with multiple sites could be foreseen to answer open questions with significant implications.

Interestingly, following the investigation of various decoherence-inducing gates,
we also extract important characters of the gate noises. For
example, we reveal that a large portion of $(X)^2$ gate errors consists
of over-rotation (see Sec. S4 in the SI). This "off-set" type of error might not as harmful as
decoherent errors to quantum computers, because it could have been
eliminated by more precise calibration of the pulses or adopting
compensating pulse sequences like our design. However, conventional
benchmark protocols only yields highly averaged metrics such as the average gate fidelity for the calibration of gate
performance, which would provide limited
information to properly describe the composition of the gate
noises. Thus, we emphasize that the detailed characterization of the
noises in current NISQ devices such as the analyses presented in the SI is still of vital importance for the
design of error-mitigation or calibration protocols for quantum simulations.   

Finally, we emphasize that the proposed decoherence-inducing gate sequences could serve as well-behaved quantum noise generator
that may also be useful for other 
applications such as state preparation or dissipative quantum
computation. We conclude that our results show the possibility of
turning NISQ devices into extremely useful and programmable platforms
to study open quantum systems under complicated system-bath
interactions, and hopefully, can help us investigate complex dynamics in open quantum
systems beyond what classical computers can do for us.

\section{Acknowledgments}
\begin{acknowledgments}
We thank IBM and IBM-Q Hub at NTU for accessing quantum devices through IBM Quantum Experience. 
YCC thanks the Ministry of Science and Technology, Taiwan (Grant 
No. MOST 109-2113-M-002-004 and MOST 109-2113-M-001-040), Physics Division, 
National Center for Theoretical Sciences (Grant No. 110-2124-M-002-012), and National 
Taiwan University (Grant No. 109L892005) for financial support. 
\end{acknowledgments}

\appendix 

\section{Experimental Methods}\label{sec:exp_methods}

Details on the encoding and propagation circuits and experimental
setup on the IBM Q experience cloud service are provided as follows.

\subsection{Encoding and Propagating the Exciton Hamiltonian}
\begin{figure}
	\centering 
	\includegraphics[width=0.8\columnwidth]{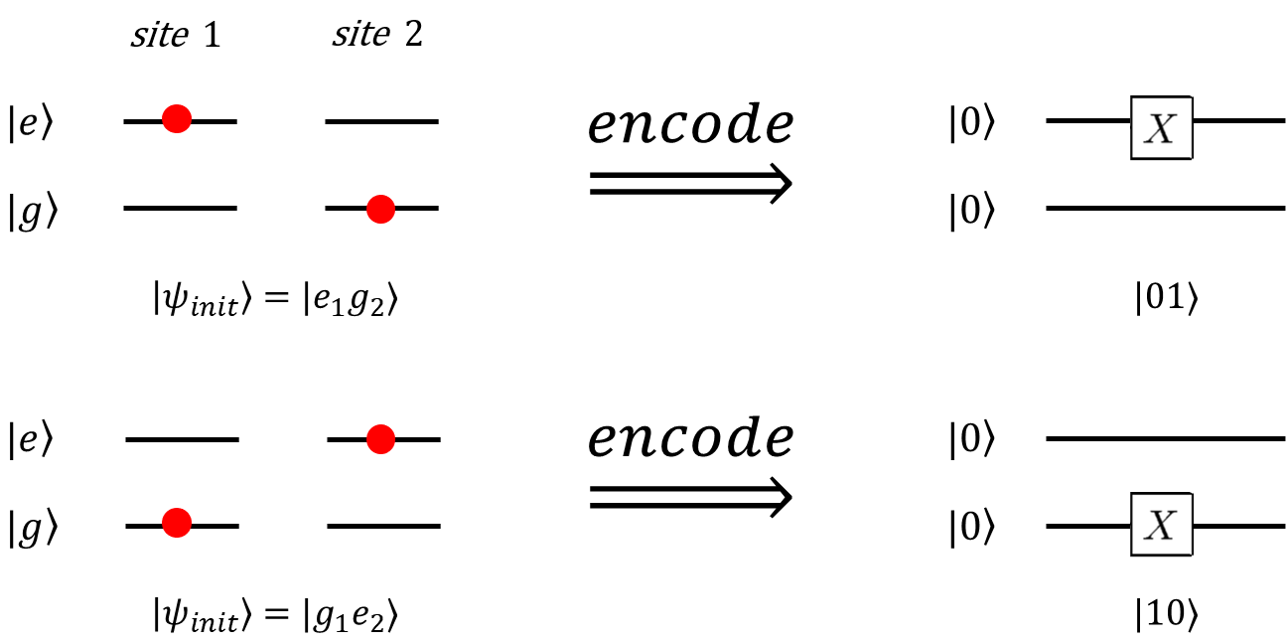}
	\caption{\textbf{Encoding the exciton states in site basis on quantum computers}
		The preparation of initial state as $|1\rangle$ or $|2\rangle$ is implemented by applying an X gate to
		the corresponding qubit.}
		\label{fig:encoding}
\end{figure}

To encode the Hamiltonian of the exciton dimer system in a quantum
computer, a Jordan-Wigner type transformation\cite{seeley2012bravyi} is
used. There, the preparation step is trivial as the state preparation circuit for exciton state with site 1 or 2 being excited can be easily implemented as shown in Fig.~\ref{fig:encoding}. Regarding mapping of the Hamiltonian, the second-quantized operators in (Eq. \ref{exciton_def}) is 
mapped to the Pauli operators acting locally on qubits according to (Eq. \ref{eq:jw_encoding})

\begin{align} \label{eq:jw_encoding}
	{a_n}^\dagger{a_n}&=\frac{1}{2}(I_n-Z_n), \nonumber \\ 
	{a_n}^\dagger{a_m}+{a_m}^\dagger{a_n}&=\frac{1}{2}(X_mX_n+Y_mY_n),\ \forall m\neq n
\end{align}
In our simulation, we model a symmetric dimer with
$J_{12}=J_0$, therefore we can encode the unit-less $J=1$ case without loss of generality. In this case, the
simulation time has a derived unit of $1/J_0$. The qubit
Hamiltonian can then be represented as 
\begin{align}
	H=\frac{1}{2}(X_1X_2+Y_1Y_2)
\end{align}
To implement the time evolution operator $e^{-iHt}$ on quantum
computers $(\hbar=1)$, one can concatenate propagators for the local
term. (Fig.~\ref{fig:quantum_circuit}(a)). 
The rotation angle $\theta$ of the phase gate $R_z(\theta)$ corresponds to the
simulation time. In all our simulation of population dynamics, we set
a single time step to be $\delta\theta=0.03767$. (This corresponds to
a step of 2 fs for a symmetric dimer system with excitonic coupling =
$100cm^{-1}$). The decoherence period ($\Delta T_D$) 
is chosen to be $25\delta\theta$, i.e. when d=1, an identity gate sequence is
prepended in the dissipation part for each 25 time steps.

\subsection{IBM Quantum Experience}
All the experiments are performed on superconducting quantum computers provided by IBM quantum
experience\cite{IBMQuantum} via qiskit framework\cite{Qiskit}. Each
job for the simulation of population dynamics consists of 150 circuits
corresponding to evenly spaced simulation time. The logical-to-physical qubit mapping
layout is chosen so that the CNOT interaction can be directly
implemented. The qubit bit string is in
little-endian format in accordance with the qiskit convention. 

\subsection{Population Dynamics from Projective Measurements}
To obtain the population in the site basis, we perform
projective measurements up to 8192 times in qubit computational basis
at each simulation time step, and calculate the population according
to $P_1(t) = \frac{N_{01}(t)}{N_{01}(t)+N_{10}(t)}$. where $N_{01}$, and $N_{10}$ are counts for the
states collapsed to $|01\rangle$, $|10\rangle$, respectively.
The counts for states collapsed to $|00\rangle$, $|11\rangle$ are excluded
because they are generated by the leakage of population to space outside of the one-exciton
manifold.




\begin{thebibliography}{10}

\bibitem{nielsen2010quantum}
Michael~A Nielsen and Isaac Chuang.
\newblock {\em Quantum computation and quantum information}.
\newblock Cambridge University Press, 2010.

\bibitem{bacon2001universal}
Dave Bacon, Andrew~M Childs, Isaac~L Chuang, Julia Kempe, Debbie~W Leung, and
  Xinlan Zhou.
\newblock Universal simulation of markovian quantum dynamics.
\newblock {\em Phys. Rev. A}, 64:062302, 2001.

\bibitem{georgescu2014quantum}
Iulia~M Georgescu, Sahel Ashhab, and Franco Nori.
\newblock Quantum simulation.
\newblock {\em Rev. Mod. Phys.}, 86:153, 2014.

\bibitem{altman2021quantum}
Ehud Altman, Kenneth~R Brown, Giuseppe Carleo, Lincoln~D Carr, Eugene Demler,
  Cheng Chin, Brian DeMarco, Sophia~E Economou, Mark~A Eriksson, Kai-Mei~C Fu,
  et~al.
\newblock Quantum simulators: Architectures and opportunities.
\newblock {\em PRX Quantum}, 2:017003, 2021.

\bibitem{yung2014introduction}
Man-Hong Yung, James~D Whitfield, Sergio Boixo, David~G Tempel, and Alan
  Aspuru-Guzik.
\newblock Introduction to quantum algorithms for physics and chemistry.
\newblock {\em Adv. Chem. Phys.}, 154:67--106, 2014.

\bibitem{OMalley:2016dc}
P~J~J O’Malley, R~Babbush, I~D Kivlichan, J~Romero, J~R McClean, R~Barends,
  J~Kelly, P~Roushan, A~Tranter, N~Ding, B~Campbell, Y~Chen, Z~Chen, B~Chiaro,
  A~Dunsworth, A~G Fowler, E~Jeffrey, E~Lucero, A~Megrant, J~Y Mutus, M~Neeley,
  C~Neill, C~Quintana, D~Sank, A~Vainsencher, J~Wenner, T~C White, P~V Coveney,
  P~J Love, H~Neven, A~Aspuru-Guzik, and J~M Martinis.
\newblock Scalable quantum simulation of molecular energies.
\newblock {\em Phys. Rev. X}, 6(3):361 -- 13, 2016.

\bibitem{Colless:2018hp}
J~I Colless, V~V Ramasesh, D~Dahlen, M~S Blok, M~E Kimchi-Schwartz, J~R
  McClean, J~Carter, W~A~de Jong, and I~Siddiqi.
\newblock Computation of molecular spectra on a quantum processor with an
  error-resilient algorithm.
\newblock {\em Phys. Rev. X}, 8(1):011021, 2018.

\bibitem{cao2019quantum}
Yudong Cao, Jonathan Romero, Jonathan~P Olson, Matthias Degroote, Peter~D
  Johnson, M{\'a}ria Kieferov{\'a}, Ian~D Kivlichan, Tim Menke, Borja
  Peropadre, Nicolas~PD Sawaya, et~al.
\newblock Quantum chemistry in the age of quantum computing.
\newblock {\em Chem. Rev.}, 119:10856--10915, 2019.

\bibitem{McArdle:2020jl}
Sam McArdle, Suguru Endo, Alan Aspuru-Guzik, Simon~C Benjamin, and Xiao Yuan.
\newblock {Quantum computational chemistry}.
\newblock {\em Rev. Mod. Phys.}, 92:015003, 2020.

\bibitem{preskill2018quantum}
John Preskill.
\newblock Quantum computing in the {NISQ} era and beyond.
\newblock {\em Quantum}, 2:79, 2018.

\bibitem{bharti2021noisy}
Kishor Bharti, Alba Cervera-Lierta, Thi~Ha Kyaw, Tobias Haug, Sumner
  Alperin-Lea, Abhinav Anand, Matthias Degroote, Hermanni Heimonen, Jakob~S.
  Kottmann, Tim Menke, Wai-Keong Mok, Sukin Sim, Leong-Chuan Kwek, and Alán
  Aspuru-Guzik.
\newblock Noisy intermediate-scale quantum ({NISQ}) algorithms.
\newblock arXiv:2101.08448v1 [quant-ph], 2021.

\bibitem{knill1998resilient}
Emanuel Knill, Raymond Laflamme, and Wojciech~H Zurek.
\newblock Resilient quantum computation.
\newblock {\em Science}, 279:342--345, 1998.

\bibitem{bravyi2018correcting}
Sergey Bravyi, Matthias Englbrecht, Robert K{\"o}nig, and Nolan Peard.
\newblock Correcting coherent errors with surface codes.
\newblock {\em npj Quantum Inf.}, 4:1--6, 2018.

\bibitem{breuer2002theory}
Heinz-Peter Breuer, Francesco Petruccione, et~al.
\newblock {\em The theory of open quantum systems}.
\newblock Oxford University Press on Demand, 2002.

\bibitem{weiss2012quantum}
Ulrich Weiss.
\newblock {\em Quantum dissipative systems}.
\newblock World Scientific, Singapore, 2012.

\bibitem{Croce:2020hs}
Roberta Croce and Herbert~van Amerongen.
\newblock {Light harvesting in oxygenic photosynthesis: Structural biology
  meets spectroscopy}.
\newblock {\em Science}, 369:933, 2020.

\bibitem{Yoneda:2020dh}
Eric~A Arsenault, Yusuke Yoneda, Masakazu Iwai, Krishna~K Niyogi, and Graham~R
  Fleming.
\newblock {Vibronic mixing enables ultrafast energy flow in light-harvesting
  complex II}.
\newblock {\em Nat. Commun.}, 11:1460, 2020.

\bibitem{Wang:2019ge}
Lili Wang, Marco~A Allodi, and Gregory~S Engel.
\newblock {Quantum coherences reveal excited-state dynamics in biophysical
  systems}.
\newblock {\em Nat. Rev. Chem.}, 3:477 -- 490, 2019.

\bibitem{Rafiq:2019jw}
Shahnawaz Rafiq and Gregory~D Scholes.
\newblock From fundamental theories to quantum coherences in electron transfer.
\newblock {\em J. Am. Chem. Soc.}, 141:708 -- 722, 2019.

\bibitem{Kim:2019ky}
Tae~Wu Kim, Sunhong Jun, Yoonhoo Ha, Rajesh~K. Yadav, Abhishek Kumar, Chung-Yul
  Yoo, Inhwan Oh, Hyung-Kyu Lim, Jae~Won Shin, Ryong Ryoo, Hyungjun Kim,
  Jeongho Kim, Jin-Ook Baeg, and Hyotcherl Ihee.
\newblock {Ultrafast charge transfer coupled with lattice phonons in
  two-dimensional covalent organic frameworks}.
\newblock {\em Nat. Commun.}, 10:1873, 2019.

\bibitem{zurek2003decoherence}
Wojciech~Hubert Zurek.
\newblock Decoherence, einselection, and the quantum origins of the classical.
\newblock {\em Rev. Mod. Phys.}, 75:715, 2003.

\bibitem{Tanimura:1989ab}
Yoshitaka Tanimura and Ryogo Kubo.
\newblock {Time evolution of a quantum system in contact with a nearly
  Gaussian-Markoffian noise bath}.
\newblock {\em J. Phys. Soc. Jpn.}, 58:101, 1989.

\bibitem{Tanimura:2006ab}
Yoshitaka Tanimura.
\newblock Stochastic {L}iouville, {L}angevin, {F}okker--{P}lanck, and master
  equation approaches to quantum dissipative systems.
\newblock {\em J. Phys. Soc. Jpn.}, 75:082001, 2006.

\bibitem{Ishizaki:2009ab}
Akihito Ishizaki and Graham~R Fleming.
\newblock {Unified Treatment of Quantum Coherent and Incoherent Hopping
  Dynamics in Electronic Energy Transfer: Reduced Hierarchy Equation Approach}.
\newblock {\em J. Chem. Phys.}, 130:234111, 2009.

\bibitem{Jin:2008ab}
Jinshuang Jin, Xiao Zheng, and YiJing Yan.
\newblock {Exact Dynamics of Dissipative Electronic Systems and Quantum
  Transport: Hierarchical Equations of Motion Approach}.
\newblock {\em J. Chem. Phys.}, 128:234703, 2008.

\bibitem{maniscalco2004simulating}
Sabrina Maniscalco, Jyrki Piilo, F~Intravaia, F~Petruccione, and A~Messina.
\newblock Simulating quantum {B}rownian motion with single trapped ions.
\newblock {\em Phys. Rev. A}, 69:052101, 2004.

\bibitem{chiuri2012linear}
Andrea Chiuri, Chiara Greganti, Laura Mazzola, Mauro Paternostro, and Paolo
  Mataloni.
\newblock Linear optics simulation of quantum non-markovian dynamics.
\newblock {\em Sci. Rep.}, 2:1--5, 2012.

\bibitem{mostame2012quantum}
Sarah Mostame, Patrick Rebentrost, Alexander Eisfeld, Andrew~J Kerman,
  Dimitris~I Tsomokos, and Al{\'a}n Aspuru-Guzik.
\newblock Quantum simulator of an open quantum system using superconducting
  qubits: exciton transport in photosynthetic complexes.
\newblock {\em New J. Phys.}, 14:105013, 2012.

\bibitem{potovcnik2018studying}
Anton Poto{\v{c}}nik, Arno Bargerbos, Florian~AYN Schr{\"o}der, Saeed~A Khan,
  Michele~C Collodo, Simone Gasparinetti, Yves Salath{\'e}, Celestino Creatore,
  Christopher Eichler, Hakan~E T{\"u}reci, et~al.
\newblock Studying light-harvesting models with superconducting circuits.
\newblock {\em Nat. Commun.}, 9:1--7, 2018.

\bibitem{wang2018efficient}
Bi-Xue Wang, Ming-Jie Tao, Qing Ai, Tao Xin, Neill Lambert, Dong Ruan,
  Yuan-Chung Cheng, Franco Nori, Fu-Guo Deng, and Gui-Lu Long.
\newblock Efficient quantum simulation of photosynthetic light harvesting.
\newblock {\em npj Quantum Inf.}, 4:1--6, 2018.

\bibitem{trautmann2018trapped}
Nils Trautmann and Philipp Hauke.
\newblock Trapped-ion quantum simulation of excitation transport: Disordered,
  noisy, and long-range connected quantum networks.
\newblock {\em Phys. Rev. A}, 97:023606, 2018.

\bibitem{maier2019environment}
Christine Maier, Tiff Brydges, Petar Jurcevic, Nils Trautmann, Cornelius
  Hempel, Ben~P Lanyon, Philipp Hauke, Rainer Blatt, and Christian~F Roos.
\newblock Environment-assisted quantum transport in a 10-qubit network.
\newblock {\em Phys. Rev. Lett.}, 122:050501, 2019.

\bibitem{su2020quantum}
Hong-Yi Su and Ying Li.
\newblock Quantum algorithm for the simulation of open-system dynamics and
  thermalization.
\newblock {\em Phys. Rev. A}, 101:012328, 2020.

\bibitem{garcia2020ibm}
Guillermo Garc{\'\i}a-P{\'e}rez, Matteo~AC Rossi, and Sabrina Maniscalco.
\newblock Ibm q experience as a versatile experimental testbed for simulating
  open quantum systems.
\newblock {\em npj Quantum Inf.}, 6:1--10, 2020.

\bibitem{rost2020simulation}
Brian Rost, Barbara Jones, Mariya Vyushkova, Aaila Ali, Charlotte Cullip,
  Alexander Vyushkov, and Jarek Nabrzyski.
\newblock Simulation of thermal relaxation in spin chemistry systems on a
  quantum computer using inherent qubit decoherence.
\newblock {\em arXiv:2001.00794}, 2020.

\bibitem{feynman2018simulating}
Richard~P Feynman.
\newblock Simulating physics with computers.
\newblock {\em Int. J. Theor. Phys}, 21:467--488, 1982.

\bibitem{lloyd1996universal}
Seth Lloyd.
\newblock Universal quantum simulators.
\newblock {\em Science}, pages 1073--1078, 1996.

\bibitem{klimov2018fluctuations}
PV~Klimov, Julian Kelly, Z~Chen, Matthew Neeley, Anthony Megrant, Brian
  Burkett, Rami Barends, Kunal Arya, Ben Chiaro, Yu~Chen, et~al.
\newblock Fluctuations of energy-relaxation times in superconducting qubits.
\newblock {\em Phys. Rev. Lett.}, 121:090502, 2018.

\bibitem{burnett2019decoherence}
Jonathan~J Burnett, Andreas Bengtsson, Marco Scigliuzzo, David Niepce, Marina
  Kudra, Per Delsing, and Jonas Bylander.
\newblock Decoherence benchmarking of superconducting qubits.
\newblock {\em npj Quantum Inf.}, 5:1--8, 2019.

\bibitem{lieb1961two}
Elliott Lieb, Theodore Schultz, and Daniel Mattis.
\newblock Two soluble models of an antiferromagnetic chain.
\newblock {\em Ann. Phys.}, 16:407--466, 1961.

\bibitem{low2017optimal}
Guang~Hao Low and Isaac~L Chuang.
\newblock Optimal hamiltonian simulation by quantum signal processing.
\newblock {\em Phys. Rev. Lett.}, 118:010501, 2017.

\bibitem{IBMQuantum}
{IBM} quantum experience.
\newblock \url{https://quantum-computing.ibm.com/}.

\bibitem{mckay2017efficient}
David~C McKay, Christopher~J Wood, Sarah Sheldon, Jerry~M Chow, and Jay~M
  Gambetta.
\newblock Efficient {Z} gates for quantum computing.
\newblock {\em Phys. Rev. A}, 96:022330, 2017.

\bibitem{merrill2014progress}
J~True Merrill and Kenneth~R Brown.
\newblock Progress in compensating pulse sequences for quantum computation.
\newblock {\em Adv. Phys. Chem.}, pages 241--294, 2014.

\bibitem{barreiro2011open}
Julio~T Barreiro, Markus M{\"u}ller, Philipp Schindler, Daniel Nigg, Thomas
  Monz, Michael Chwalla, Markus Hennrich, Christian~F Roos, Peter Zoller, and
  Rainer Blatt.
\newblock An open-system quantum simulator with trapped ions.
\newblock {\em Nature}, 470:486--491, 2011.

\bibitem{Mohseni:2008p67347}
Masoud Mohseni, Patrick Rebentrost, Seth Lloyd, and Alan Aspuru-Guzik.
\newblock {Environment-assisted quantum walks in photosynthetic energy
  transfer}.
\newblock {\em J. Chem. Phys.}, 129:174106, 2008.

\bibitem{Chang:2012il}
Hung-Tzu Chang and Yuan-Chung Cheng.
\newblock {Coherent versus incoherent excitation energy transfer in molecular
  systems}.
\newblock {\em J. Chem. Phys.}, 137:165103, 2012.

\bibitem{strumpfer2012open}
Johan Strümpfer and Klaus Schulten.
\newblock Open quantum dynamics calculations with the hierarchy equations of
  motion on parallel computers.
\newblock {\em J. Chem. Theory Comput.}, 8:2808--2816, 2012.

\bibitem{cerrillo2014non}
Javier Cerrillo and Jianshu Cao.
\newblock Non-{M}arkovian dynamical maps: numerical processing of open quantum
  trajectories.
\newblock {\em Phys. Rev. Lett.}, 112:110401, 2014.

\bibitem{mohseni2008quantum}
Masoud Mohseni, AT~Rezakhani, and DA~Lidar.
\newblock Quantum-process tomography: Resource analysis of different
  strategies.
\newblock {\em Phys. Rev. A}, 77:032322, 2008.

\bibitem{gross2010quantum}
David Gross, Yi-Kai Liu, Steven~T Flammia, Stephen Becker, and Jens Eisert.
\newblock Quantum state tomography via compressed sensing.
\newblock {\em Phys. Rev. Lett.}, 105:150401, 2010.

\bibitem{shabani2011efficient}
A~Shabani, RL~Kosut, M~Mohseni, H~Rabitz, MA~Broome, MP~Almeida, A~Fedrizzi,
  and AG~White.
\newblock Efficient measurement of quantum dynamics via compressive sensing.
\newblock {\em Phys. Rev. Lett.}, 106:100401, 2011.

\bibitem{seeley2012bravyi}
Jacob~T Seeley, Martin~J Richard, and Peter~J Love.
\newblock The {B}ravyi-{K}itaev transformation for quantum computation of
  electronic structure.
\newblock {\em J. Chem. Phys.}, 137:224109, 2012.

\bibitem{Qiskit}
H{\'e}ctor Abraham et~al.
\newblock Qiskit: An open-source framework for quantum computing.
\newblock DOI: 10.5281/zenodo.2562110, 2019.

\end{thebibliography}

\clearpage

\pagestyle{empty}	
\includegraphics[scale=1, page=1]{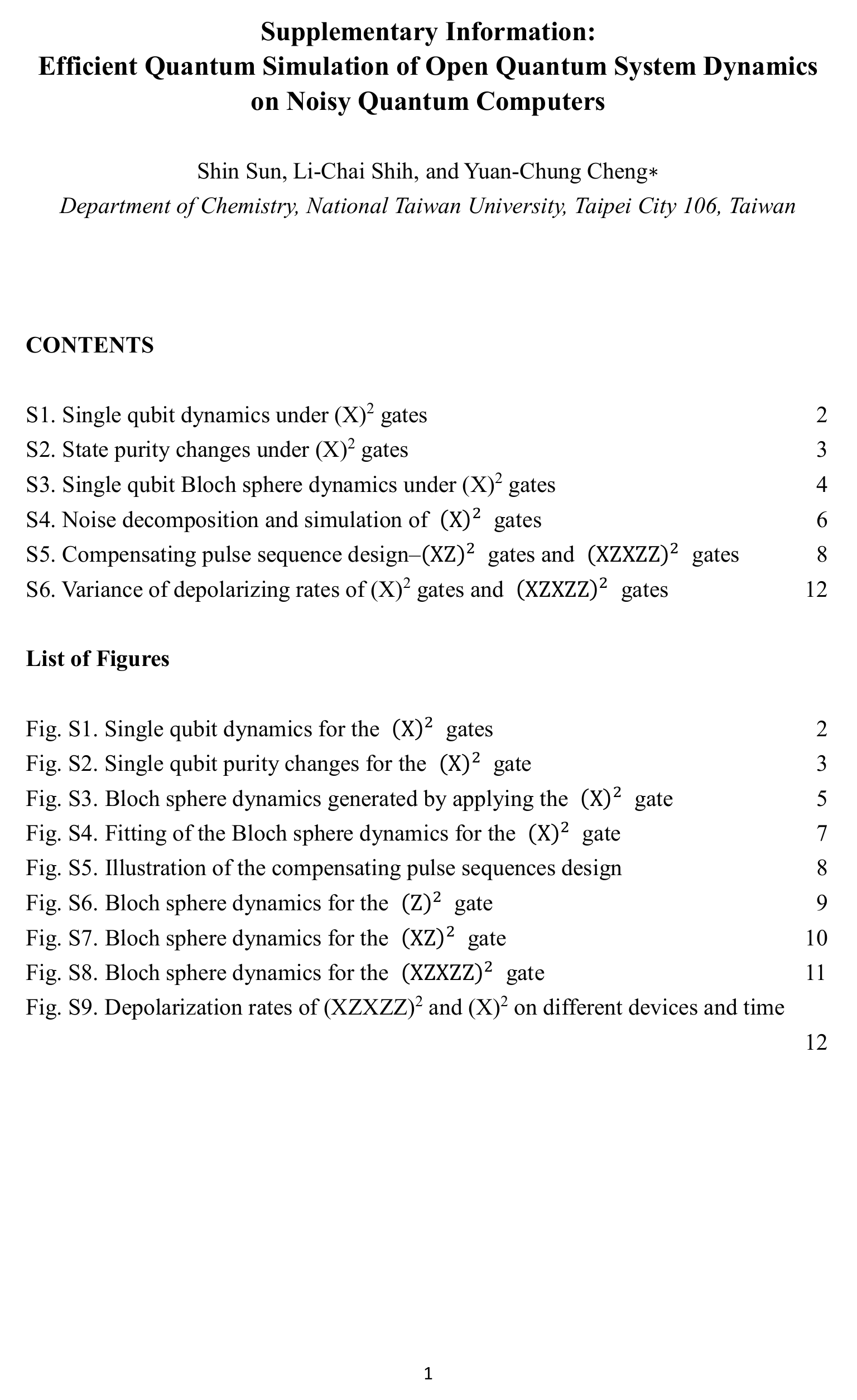}
\clearpage
\includegraphics[scale=1, page=2]{si.pdf}
\clearpage
\includegraphics[scale=1, page=3]{si.pdf}
\clearpage
\includegraphics[scale=1, page=4]{si.pdf}
\clearpage
\includegraphics[scale=1, page=5]{si.pdf}
\clearpage
\includegraphics[scale=1, page=6]{si.pdf}
\clearpage
\includegraphics[scale=1, page=7]{si.pdf}
\clearpage
\includegraphics[scale=1, page=8]{si.pdf}
\clearpage
\includegraphics[scale=1, page=9]{si.pdf}
\clearpage
\includegraphics[scale=1, page=10]{si.pdf}
\clearpage
\includegraphics[scale=1, page=11]{si.pdf}
\clearpage
\includegraphics[scale=1, page=12]{si.pdf}
\clearpage
\end{document}